\documentstyle[12pt]{article}
\newcommand{\be}{\begin{equation}}
\newcommand{\ee}{\end{equation}}

\begin{document}

\begin{titlepage}
\begin{flushright}
CAMS/00-02\\
\end{flushright}
\vspace{.6cm}
\begin{center}
\renewcommand{\thefootnote}{\fnsymbol{footnote}}
{\Large \bf Dimensional Reduction, Gauged $D=5$ Supergravity and Brane Solutions}
\vfill
{\large \bf {A. H. Chamseddine \footnote{email: chams@aub.edu.lb} and
W.~A.~Sabra\footnote{email: ws00@aub.edu.lb}}}\\
\renewcommand{\thefootnote}{\arabic{footnote}}
\setcounter{footnote}{0}
\vfill
{\small
Center for Advanced Mathematical Sciences (CAMS)\\
and\\
Physics Department, American University of Beirut, Lebanon.}
\end{center}
\vfill
\begin{center}
{\bf Abstract}
\end{center}
The $U(1)$  gauged version of the Strominger-Vafa five dimensional $N=2$ supergravity
with one vector multiplet is obtained via dimensional reduction from the $N=1$ ten
dimensional supergavity. Using such explicit  relation between  the gauged supergravity
theory and ten
dimensional supergravity, all known solutions of the five dimensional theory can be  lifted up to
ten-dimensions. The eleven dimensional solutions can also obtained by lifting
the ten-dimensional solutions.
\end{titlepage}

\section{\protect\bigskip Introduction}

Recently, there has been renewed activities in the study of gauged
supergravity in various dimensions as well as in their solutions. This, to a
large extent, is motivated by the conjectured equivalence between string
theory on anti-de~Sitter (AdS) spaces (times some compact manifold) and
certain superconformal gauge theories living on the boundary of AdS \cite
{ads}. The theories of extended supergravity in various dimensions possess
rigid symmetries. A subgroup of these symmetries can be gauged by the vector
fields present in the ungauged theory. Gauged supergravity theories exist in
space-time dimensions where supersymmetry allows the existence of a
cosmological constant. In $D=11$, $D=10$ and $D=9,$ a cosmological constant
is not possible.

In five dimensions, $N=2$ supergravity theories can be obtained by gauging
the $U(1)$ subgroup of the $SU(2)$ automorphism group of the $N=2$
supersymmetry algebra, thus breaking $SU(2)$ down to the $U(1)$ group. The $%
U(1)$ gauge field introduced to gauge the theory can be taken as a linear
combination of the abelian vector fields of the ungauged theory with a
coupling constant $g$. The additional couplings of the fermi-fields of the
ungauged theory to the $U(1)$ gauge vector field breaks supersymmetry which
necessitates the addition of $g$-dependent gauge invariant terms in order to
restore $N=2$ supersymmetry. The purely bosonic terms added produce the
scalar potential of the theory \cite{gunaydin}.

Gauged supergravity can be obtained by compactifying higher dimensional
supergravity on a group manifold. It is usually difficult to find a
consistent ansatz for the compactification\footnote{%
see \cite{vaman} and references therein.}. In particular it was shown
recently \cite{CV} that the Freedman-Schwarz \cite{FS} gauged $N=4$
supergravity can be obtained by compactifying ten dimensional supergravity
on the $SU(2)\times SU(2)$ group manifold. The difficulty arises because one
has to identify the vector fields coming from the compactification of the
metric with the vector fields coming from the antisymmetric tensor. The
vector fields coming from compactifying the metric on a group manifold
behave properly as $SU(2)\times SU(2)$ gauge fields. The components of the
antisymmetric tensor do not usually behave like $SU(2)\times SU(2)$ gauge
fields, and for this to happen a very precise form for the ansatz of the
antisymmetric tensor must be taken. This prescription also works in
obtaining $D=7$ gauged supergravity by compactifying ten dimensional
supergravity on $SU(2)$ group manifold as was recently shown in \cite{csd7}.
In a related matter, there are now many known black hole solutions\ for
gauged $D=5$ supergravity \cite{e12, e13, m, SK}. These solutions could be
promoted to solutions of ten and eleven dimensional supergravity if one can
embed the five dimensional supergravity in the higher dimensional ones.

It is our purpose in this paper to show that by compactifying and truncating
$D=10$ supergravity to $D=5$ on an $SU(2)\times $ $U(1)^{2}$ group manifold
one obtains a gauged $D=5$ supegravity theory with one vector multiplet. The
solutions for this model can then be lifted to seven, ten and eleven
dimensions.\ This work is organized as follow. In section two, it is shown
how to reduce $D=10$ supergravity to a particular gauged {\it N=2} five
dimensional supergravity theory coupled to one vector multiplet. The gauged
theory obtained is the $U(1)$ gauged version of the model introduced by
Strominger and Vafa\cite{SV}. The five dimensional model obtained is then
reformulated in the framework of very special geometry in section three.
Some particular solutions of the five dimensional theory, as examples, are
lifted to seven, ten and eleven dimensions in section four. Finally our
results are summarized and discussed.

\section{A $D=5$ Gauged Supergravity From $D=10$ Supergravity}

In this section we consider the dimensional reduction of $D=10$ supergravity
down to $N=2$, $U(1)$ gauged $D=5$ supergravity coupled to one vector
multiplet. First, the bosonic part of $N=1$ supergravity action in ten
dimensions is
\begin{eqnarray}
S_{10} &=&\int \hat{e}\left( -\frac{1}{4}\,\hat{R}+\frac{1}{2}\,\partial _{M}%
\hat{\phi}\,\partial ^{M}\hat{\phi}+\frac{1}{12}\,e^{-2\hat{\phi}}\hat{H}%
_{MNP}\,\hat{H}^{MNP}\right) \,d^{4}x\,d^{6}z  \nonumber \\
&\equiv &S_{\hat{G}}+S_{\hat{\phi}}+S_{\hat{H}}.  \label{tenaction}
\end{eqnarray}

The notation used in this paper is as follows. We denote ten-dimensional
quantities by hatted symbols. Base and tangent space indices are denoted by
late and early capital Latin letters, respectively. For the four-dimensional
space-time, we use late and early Greek letters, respectively, to denote
base space and tangent space indices. Similarly, the internal base space and
tangent space indices are denoted by late and early Latin letters,
respectively.
\begin{equation}
\{M\}=\{\mu =0,\ldots ,3;\,m=1,\ldots ,6\},\ \ \ \{A\}=\{\alpha =0,\ldots
,3;\,a=1,\ldots ,6\}.\ \
\end{equation}
The general coordinates $\hat{x}^{M}$ consist of spacetime coordinates $x^{{%
\mu }}$ and internal coordinates $z^{m}$. The flat Lorentz metric of the
tangent space is chosen to be $(+,-,\ldots ,-)$ with the internal dimensions
all spacelike. Thus the metric is related to the vielbein by
\begin{equation}
\hat{{\bf g}}_{MN}=\hat{\eta}_{AB}\hat{e}_{\ M}^{A}\hat{e}_{\ N}^{B}=\eta
_{\alpha \beta }\hat{e}_{\ M}^{\alpha }\hat{e}_{\ N}^{\beta }-\delta _{ab}%
\hat{e}_{\ M}^{a}\hat{e}_{\ N}^{b}\,,
\end{equation}
and the antisymmetric tensor field strength is
\begin{equation}
\hat{H}_{MNP}=\partial _{M}\hat{B}_{NP}+\partial _{N}\hat{B}_{PM}+\partial
_{P}\hat{B}_{MN}\,.
\end{equation}
The coordinates $z^{m}$ span the internal compact group space. Thus we
introduce the functions $U_{m}^{a}(z)$ which satisfy the condition \cite{SS}
\begin{equation}
\left( U^{-1}\right) _{b}^{\ m}\left( U^{-1}\right) _{c}^{\ n}\left(
\partial _{m}U_{\ n}^{a}-\partial _{n}U_{\ m}^{a}\right) =\frac{f_{abc}}{%
\sqrt{2}}\,,
\end{equation}
Here $f_{abc}$ are the group structure constants and the internal space
volume is $\Omega =\int |U_{m}^{a}|d^{6}{\bf z}$. \ In the maximal case, i.
e., $SU(2)\times SU(2)$, each $S^{3}$ factor admits invariant 1-form $\theta
^{a}=\theta _{i}^{a}dz^{i}$, which satisfies
\begin{equation}
d\theta ^{a}+{\frac{1}{2}}\epsilon _{abc}\theta ^{b}\wedge \theta ^{c}=0.
\end{equation}
If one chooses
\begin{equation}
U_{m}^{a}\equiv U_{i}^{a}=-{\frac{\sqrt{2}}{g}}\theta _{i}^{a},
\end{equation}
where $g$ is a coupling constant, then the structure constants will be given
in terms of the coupling constant by $f_{abc}=g\epsilon _{abc}$. For the
case where the coupling constant of one of the $SU(2)$ factors vanishes, the
internal space becomes the group manifold $SU(2)\times \lbrack U(1)]^{3}$.

Our ansatz for the reduction to five dimensions is given by the following
parameterization of the vielbein
\begin{equation}
\hat{e}_{M}^{A}=\left( \matrix{e^{{5\over6}\hat\phi} e^\alpha_\mu& {\sqrt2}
A_\mu^a e^{-{1\over2}\hat\phi}\cr 0 & e^{-{1 \over2}\hat\phi} U^a_m\cr}%
\right) .  \label{eansatz}
\end{equation}
Here the function $U$ depends only on the internal coordinates $(5,\cdots
,9) $. As an internal space we take the group manifold $SU(2)\times \lbrack
U(1)]^{2}$, this means that the following choice for the structure constants
is taken,
\begin{eqnarray}
f_{mnp} &=&g\epsilon _{mnp},\qquad m,n,p=5,6,7,\qquad   \nonumber
\\ f_{mnp} &=&0,\qquad \qquad m,n,p=8,9
\end{eqnarray}
The $S_{\hat{G}}$ term of $D=10$ supergravity gives upon reduction, the
following $D=5$ Lagrangian
\begin{equation}
{\cal L}_{5G}=e\left( -{\frac{1}{4}}R-{\frac{1}{8}}e^{-{\frac{8}{3}}\hat{\phi%
}}F_{\mu \nu }^{a}F^{\mu \nu a}+{\frac{5}{6}}g^{\mu \nu }\partial _{\mu }%
\hat{\phi}\partial _{\nu }\hat{\phi}+{\frac{3g^{2}}{16}}e^{{\frac{8}{3}}\hat{%
\phi}}\right)
\end{equation}
\begin{equation}
F_{\mu \nu }^{a}=\partial _{\mu }A_{\nu }^{a}-\partial _{\nu }A_{\mu
}^{a}+f_{abc}A_{\mu }^{b}A_{\nu }^{c}
\end{equation}
This is obtained by using the general formula of Scherk and Schwarz \cite{SS}
for reducing gravity from higher dimensions. For the antisymmetric tensor,
we take the ansatz
\begin{equation}
{\hat{B}}_{\mu \nu }={B}_{\mu \nu },\quad {\hat{B}}_{\mu m}=-{\frac{1}{\sqrt{%
2}}}{A}_{\mu }^{a}U_{m}^{a}(z),\quad {\hat{B}}_{mn}={\tilde{B}}_{mn}(z)
\label{bansatz}
\end{equation}
such that $\hat{H}_{mnp}=\frac{1}{2\sqrt{2}}f_{mnp}.$ To evaluate $\hat{H}%
_{MNP}\,\hat{H}^{MNP}$ we first define
\[
\hat{H}_{ABC}=\hat{e}_{A}^{M}\hat{e}_{B}^{N}\hat{e}_{C}^{P}\hat{H}_{MNP}.
\]
A direct substitution of the ansatz $\left( \ref{eansatz}\right) $ and $%
\left( \ref{bansatz}\right) $ gives
\begin{eqnarray*}
\hat{H}_{\alpha \beta \gamma } &=&e^{-\frac{5}{2}\hat{\phi}}e_{\alpha }^{\mu
}e_{\beta }^{\nu }e_{\gamma }^{\rho }H_{\mu \nu \rho }^{\prime }, \\
\hat{H}_{\alpha \beta c} &=&\frac{1}{\sqrt{2}}e^{-\frac{7}{6}\hat{\phi}%
}e_{\alpha }^{\mu }e_{\beta }^{\nu }F_{\mu \nu }^{a}, \\
\hat{H}_{\alpha bc} &=&0, \\
\hat{H}_{abc} &=&\frac{1}{2\sqrt{2}}f_{abc}
\end{eqnarray*}
where
\begin{eqnarray}
H_{\mu \nu \rho }^{\prime } &=&H_{\mu \nu \rho }-\omega _{\mu \nu \rho },
\nonumber \\
\omega _{\mu \nu \rho } &=&-6(A_{[\mu }^{a}\partial _{\nu }A_{\rho ]}^{a}+{%
\frac{1}{3}}f_{abc}A_{\mu }^{a}A_{\nu }^{b}A_{\rho }^{c}).
\end{eqnarray}
These results are a very strong consistency checks on the ansatz, especially
in the form of $H_{\mu \nu \rho }^{\prime }$ and $\hat{H}_{\alpha \beta c}.$
We now have
\begin{equation}
{\cal L}_{5B}=e\left( {\frac{1}{12}}e^{-{\frac{16}{3}}\hat{\phi}}H_{\mu \nu
\rho }^{\prime }H^{\prime \mu \nu \rho }-{\frac{1}{8}}e^{-{\frac{8\hat{\phi}%
}{3}}}F_{\mu \nu }^{a}F^{\mu \nu a}-{\frac{g^{2}}{16}}e^{{\frac{8}{3}}\hat{%
\phi}}\right) .
\end{equation}

The scalar part of $D=10$ supergravity, gives the following contribution to
the five dimensional theory
\begin{equation}
{\cal L}_{5S}={\frac{e}{2}}\partial _{\mu }\hat{\phi}\partial ^{\mu }\hat{%
\phi}.
\end{equation}
Therefore, combining all terms, the five dimensional theory is described by
the Lagrangian
\begin{equation}
{\cal L}_{5}=e\left( -{\frac{1}{4}}R-{\frac{1}{4}}e^{-{\frac{8\hat{\phi}}{3}}%
}F_{\mu \nu }^{a}F^{\mu \nu a}+{\frac{4}{3}}g^{\mu \nu }\partial _{\mu }\hat{%
\phi}\partial _{\nu }\hat{\phi}+{\frac{1}{12}}e^{-{\frac{16\hat{\phi}}{3}}%
}H_{\mu \nu \rho }^{\prime }H^{\prime \mu \nu \rho }+{\frac{g^{2}}{8}}e^{{%
\frac{8\hat{\phi}}{3}}}\right) .  \label{5dlagrangian}
\end{equation}

Since the potential of the resulting theory depends only on one scalar
field, a consistent truncation can be achieved by setting all gauge fields
but one to zero. Therefore the index $a$ in the above Lagrangian will take
one value. In order to compare with the standard Lagrangian, we multiply our
Lagrangian by a factor of $2$. Therefore the resulting $N=2$ five
dimensional theory is described by the Lagrangian
\begin{equation}
{\cal L}_{5}=e\left( -{\frac{1}{2}}R-{\frac{1}{2}}e^{-{\frac{8\hat{\phi}}{3}}%
}F_{\mu \nu }^{1}F^{\mu \nu 1}+{\frac{8}{3}}g^{\mu \nu }\partial _{\mu }\hat{%
\phi}\partial _{\nu }\hat{\phi}+{\frac{1}{6}}e^{-{\frac{16\hat{\phi}}{3}}%
}H_{\mu \nu \rho }^{\prime }H^{\prime \mu \nu \rho }+{\frac{g^{2}}{4}}e^{{%
\frac{8\hat{\phi}}{3}}}\right) ,
\end{equation}
where $\omega _{\mu \nu \rho }=-6A_{[\mu }^{1}\partial _{\nu }A_{\rho ]}^{1}$
in $H_{\mu \nu \rho }^{\prime }.$ We now apply a duality transformation by
adding to the above Lagrangian the term
\[
-\frac{1}{3}\epsilon ^{\mu \nu \rho \sigma \kappa }\partial _{\mu }A_{\nu
}^{0}H_{\rho \sigma \kappa }.
\]
Integrating $A_{\nu }^{0}$ forces $H_{\mu \nu \rho }$ to be of the form $%
3\partial _{\left[ \mu \right. }B_{\nu \rho \left. {}\right] }$ giving the
five dimensional theory with the $B_{\mu \nu }$ field. On the other hand by
integrating the independent field $H_{\mu \nu \rho }$ in the path integral
as it appears linearly and quadratically is equivalent to the substitution
\[
H_{\mu \nu \rho }=\omega _{\mu \nu \rho }+e^{\frac{16}{3}\hat{\phi}}\epsilon
_{\mu \nu \rho }^{\quad \sigma \kappa }\partial _{\sigma }A_{\kappa }^{0}.
\]
This gives the dual Lagrangian

\begin{eqnarray}
{\cal L}_{5} &=&e\left( -{\frac{1}{2}}R-{\frac{1}{2}}e^{-{\frac{8\hat{\phi}}{%
3}}}F_{\mu \nu }^{1}F^{\mu \nu 1}-{\frac{1}{2}}e^{{\frac{16\hat{\phi}}{3}}%
}F_{\mu \nu }^{0}F^{\mu \nu 0}+{\frac{8}{3}}g^{\mu \nu }\partial _{\mu }\hat{%
\phi}\partial _{\nu }\hat{\phi}\right.  \nonumber \\
&&\left. +{\frac{e^{-1}}{2}}\epsilon ^{\mu \nu \rho \sigma \lambda }F_{\mu
\nu }^{1}F_{\rho \sigma }^{1}A_{\lambda }^{0}+{\frac{g^{2}}{4}}e^{{\frac{8%
\hat{\phi}}{3}}}\right) .  \label{ra}
\end{eqnarray}
where $0$ and $1$ are to label the graviphoton and the additional vector
multiplet gauge fields respectively. \ This is the gauged $U(1)$ five
dimensional $N=2$ supergravity with one vector multiplet. It is the gauged
version of the five dimensional theory initially introduced by Strominger
and Vafa \cite{SV}.

We close this section by noting that it is possible to obtain $D=5$ gauged
supergravity from $D=7$ gauged supergravity. In a recent work we have
derived $D=7$ gauged supergravity by compactifying $D=10$ on an $SU(2)$
group manifold \cite{csd7}. The Lagrangian in seven dimensions is given by
\[
{\cal L}_{7}=e\left( -{\frac{1}{4}}R-{\frac{1}{4}}e^{-{\frac{8}{5}}\hat{\phi}%
}F_{MN}^{a}F^{MNa}+{\frac{4}{5}}g^{MN}\partial _{M}\hat{\phi}\partial _{N}%
\hat{\phi}+{\frac{1}{12}}e^{-{\frac{16}{5}}\hat{\phi}}H_{MNP}^{\prime
}H^{^{\prime }MNP}+{\frac{g^{2}}{8}}e^{{\frac{8}{5}}\hat{\phi}}\right) .
\]
This is reduced and truncated to $D=5$ gauged supergravity by taking the
following ansatz
\[
e_{M}^{A}=\left(
\begin{array}{cc}
e^{\frac{8}{15}\hat{\phi}}e_{\mu }^{\alpha } & 0 \\
0 & e^{-\frac{4}{5}\hat{\phi}}\delta _{m}^{i}
\end{array}
\right) ,\quad m=5,6.
\]
as well as $B_{\mu m}=0$ and $A_{m}^{a}=0$. One can easily show that the
reduced Lagrangian is given by $\left( \ref{5dlagrangian}\right) $

\section{\protect\bigskip Embedding Into Very Special Geometry}

The solutions of five dimensional $N=2$ supergravity theory with vector
multiplets theory have been discussed within the framework of very special
geometry \cite{e12, e13, m, SK}. Therefore, before we discuss solutions of
the five dimensional theory and their embedding into ten dimensional
supergravity and M-theory, it is essential to consider our compactified
Lagrangian in this framework.

A class of five-dimensional $N=2$ supergravity coupled to abelian vector
supermultiplets can be obtained by compactifying eleven-dimensional
supergravity, the low-energy theory of M-theory, on a Calabi-Yau three-folds
\cite{townsend}. The massless spectrum of the theory contains $(h_{(1,1)}-1)$
vector multiplets with real scalar components, and thus $h_{(1,1)}$ vector
bosons (the additional vector boson is the graviphoton). The theory also
contains $h_{(2,1)}+1$ hypermultiplets, where $h_{(1,1)}$ and $h_{(2,1)},$
are the Calabi-Yau Hodge numbers.

The bosonic part of the effective gauged supersymmetric $N=2$ Lagrangian
which describes the coupling of vector multiplets to supergravity is given
by \cite{gunaydin}
\begin{equation}
{\cal L}=e\left( {\frac{1}{2}}R+g^{2}V-{\frac{1}{4}}G_{IJ}F_{\mu \nu
}{}^{I}F^{\mu \nu J}-{\frac{1}{2}}{\cal G}_{ij}\partial _{\mu }\phi
^{i}\partial ^{\mu }\phi ^{j}+{\frac{e^{-1}}{48}}\epsilon ^{\mu \nu \rho
\sigma \lambda }C_{IJK}F_{\mu \nu }^{I}F_{\rho \sigma }^{J}A_{\lambda
}^{K}\,\right) ,  \label{action}
\end{equation}
$R$ is the scalar curvature, $F_{\mu \nu }^{I}$ are the Abelian
field-strength tensor, $V$ is the potential given by
\begin{equation}
V(X)=V_{I}V_{J}\left( 6X^{I}X^{J}-{\frac{9}{2}}{\cal G}^{ij}\partial
_{i}X^{I}\partial _{j}X^{J}\right) ,
\end{equation}
where $X^{I}$ represent the real scalar fields which have to satisfy the
constraint
\begin{equation}
{\cal V}={\frac{1}{6}}C_{IJK}X^{I}X^{J}X^{K}=1\ .
\end{equation}
Also:
\begin{equation}
G_{IJ}=-{\frac{1}{2}}\partial _{I}\partial _{J}\log {\cal V}\Big|_{{\cal V}%
=1}\quad ,\quad {\cal G}_{ij}=\partial _{i}X^{I}\partial _{j}X^{J}G_{IJ}\Big|%
_{{\cal V}=1}\ ,
\end{equation}
where $\partial _{i}$ refers to a partial derivative with respect to the
scalar field $\phi ^{i}$. The physical quantities in (\ref{action}) can all
be expressed in terms of the homogeneous cubic polynomial ${\cal V}$.

Further useful relations are
\begin{equation}
\partial _{i}X_{I}=-{\frac{2}{3}}G_{IJ}\partial _{i}X^{J}\quad ,\quad X_{I}={%
\frac{2}{3}}G_{IJ}X^{J}\ .  \label{alsouseful}
\end{equation}
It is worth pointing out that for Calabi-Yau compactification, ${\cal V}$ is
the intersection form, $X^{I}$ and $X_{I}={\frac{1}{6}}C_{IJK}X^{J}X^{K}$
correspond to the size of the two- and four-cycles and $C_{IJK}$ are the
intersection numbers of the Calabi-Yau threefold.

A very useful relation of very special geometry is
\begin{equation}
{\cal G}^{ij}\partial _{j}X^{I}\partial _{j}X^{J}=G^{IJ}-{\frac{2}{3}}%
X^{I}X^{J}\ .  \label{useful}
\end{equation}
The potential can also be written as
\begin{equation}
V(X)=9\,V_{I}V_{J}\left( X^{I}X^{J}-{\frac{1}{2}}G^{IJ}\right) \ .
\end{equation}

The Lagrangian (\ref{ra}) correspond to the following identifications%
\footnote{%
note that the sign difference of the kinetic terms is due to using a
different metric signature}
\[
G_{00}=2e^{\frac{16\hat{\phi}}{3}},\qquad G_{11}=2e^{-8{\frac{\hat{\phi}}{3}}%
},{\cal G}_{11}=\frac{16}{3}, \quad C_{011}=8.
\]

In order to determine the $X^{I},$ we use the relations (\ref{alsouseful})
together with $X_{I}X^{I}=1,$ this gives $X^{0}=c_{1}e^{-8{\frac{\hat{\phi}}{%
3}}}$, $X^{1}=c_{2}e^{4{\frac{\hat{\phi}}{3}}}.$ Upon using (\ref{useful})
and the above identification we get $c_{1}=\frac{1}{2},$ $c_{2}=\frac{1}{%
\sqrt{2}}.$ Finally, from the expression of the potential one obtains that $%
V_{0}=0,$ $V_{1}=\frac{1}{3}.$ Therefore we have

\[
X^{0}=\frac{1}{2}e^{-8{\frac{\hat{\phi}}{3}}},\qquad %
X^{1}=\frac{1}{\sqrt{2}}e^{4{\frac{\hat{\phi}}{3}}},\qquad %
X_{0}=\frac{2}{3}e^{8{\frac{\hat{\phi}}{3}}},\qquad X_{1}=%
\frac{2\sqrt{2}}{3}e^{-4{\frac{\hat{\phi}}{3}}}.
\]

\section{Lifting Solutions to ten and eleven dimensions}

Our previous results suggest that any solution of gauged supergravity in
seven and five dimension given in terms of the metric, gauge field and
scalar fields can be lifted to ten dimensions as a solution of $N=1$ ten
dimensional supergravity. By also noting the relation between the ten and
eleven dimensional theory, one can then lift all solutions to eleven
dimensions.

To lift the solutions to ten dimensions we have to express the ten
dimensional fields in terms of five dimensional ones. To start with we write
the ten dimensional metric as
\begin{eqnarray*}
\hat{g}_{\mu \nu } &=&e^{\frac{5}{3}\hat{\phi}}g_{\mu \nu }-2e^{-\hat{\phi}%
}A_{\mu }^{a}A_{\nu }^{a}, \\
\hat{g}_{\mu m} &=&-\sqrt{2}e^{-\hat{\phi}}A_{\mu }^{a}U_{m}^{a}\left(
z\right) , \\
\hat{g}_{mn} &=&-e^{-\hat{\phi}}U_{m}^{a}\left( z\right) U_{n}^{a}\left(
z\right) .
\end{eqnarray*}
The five dimensional model is given in terms of $X^{0},X^{1},A_{\mu }^{0}$
and $A_{\mu }^{1}.$ Since $A_{\mu }^{0}$ is the dual of $B_{\mu \nu }$ we
first evaluate
\[
H_{\mu \nu \rho }^{\prime }=H_{\mu \nu \rho }-\omega _{\mu \nu \rho }=e^{%
\frac{16}{3}\hat{\phi}}\epsilon _{\mu \nu \rho }^{\quad \sigma \kappa
}\partial _{\sigma }A_{\kappa }^{0}.
\]
Or in terms of the ten dimensional fields
\begin{eqnarray*}
\hat{H}_{\alpha \beta \gamma } &=&e^{-\frac{5}{2}\hat{\phi}}e_{\alpha }^{\mu
}e_{\beta }^{\nu }e_{\gamma }^{\rho }H_{\mu \nu \rho }^{\prime }=e^{\frac{17%
}{6}\hat{\phi}}\epsilon _{\alpha \beta \gamma }^{\quad \delta \eta
}e_{\delta }^{\sigma }e_{\eta }^{\kappa }\partial _{\sigma }A_{\kappa }^{0},
\\
\hat{H}_{\alpha \beta c} &=&\frac{1}{\sqrt{2}}\delta _{c}^{1}e^{-\frac{7}{6}%
\hat{\phi}}e_{\alpha }^{\mu }e_{\beta }^{\nu }F_{\mu \nu }^{1}.
\end{eqnarray*}
The dilaton field is determined from the solution to $X^{0}$ and $X^{1}.$

\subsection{\protect\bigskip Electrically charged solutions}

The spherically symmetric BPS electric solutions as well as magnetic string
solutions were obtained in \cite{e12, m} by solving for the vanishing of the
gravitino and gaugino supersymmetry variation for a particular choice of the
supersymmetry parameter. These are given by
\begin{eqnarray}
ds^{2} &=&-{\cal {V}}^{-4/3}(1+g^{2}r^{2}{\cal V}^{2})dt^{2}+{\cal V}^{2/3}%
\left[ {\frac{dr^{2}}{1+g^{2}r^{2}{\cal V}^{2}}}+r^{2}(d\theta ^{2}+\sin
^{2}\theta d\phi ^{2}+\cos ^{2}\theta d\psi ^{2})\right]  \nonumber \\
F_{tm}^{I} &=&-\partial _{m}({\cal V}^{-1}Y^{I})\qquad ,\qquad  \nonumber \\
{\cal V} &=&{\frac{1}{6}}C_{IJK}Y^{I}Y^{J}Y^{K},\qquad {\frac{1}{2}}%
C_{IJK}Y^{J}Y^{K}=H_{I}=3V_{I}+{\frac{q_{I}}{r^{2}}}  \label{solution}
\end{eqnarray}
where
\[
Y^{I}={\cal V}^{\frac{2}{3}}X^{I},\quad Y_{I}={\cal V}^{\frac{1}{3}%
}X_{I},\quad {\cal V}=e^{3U}
\]

\bigskip Let us go back to the solutions of our five dimensional gauged
theory with one vector multiplet. As for the electric solutions, we have the
following equations
\begin{eqnarray}
{\cal V} &=&{\frac{1}{2}}C_{011}Y^{0}Y^{1}Y^{1}=4Y^{0}(Y^{1})^{2},  \nonumber
\\
&&{\frac{1}{2}}C_{011}(Y^{1})^{2}=4(Y^{1})^{2}=H_{0}={\frac{q_{0}}{r^{2}},}
\nonumber \\
&&C_{110}Y^{1}Y^{0}=8(Y^{1})Y^{0}=H_{1}=1+{\frac{q_{1}}{r^{2}}.}
\label{ssolution}
\end{eqnarray}
This gives
\begin{eqnarray*}
Y^{0} &=&\frac{r}{4q_{0}^{\frac{1}{2}}}\left( 1+\frac{q_{1}}{r^{2}}\right) ,
\\
Y^{1} &=&\frac{q_{0}^{\frac{1}{2}}}{2r}, \\
{\cal V} &=&\frac{q_{0}^{\frac{1}{2}}}{4r}\left( 1+\frac{q_{1}}{r^{2}}%
\right) .
\end{eqnarray*}
The metric depends only on ${\cal V}$. The gauge field strengths are given by

\begin{eqnarray}
F_{tr}^{0} &=&-\partial _{r}({\cal V}^{-1}Y^{0})=-\frac{2r}{q_{0}}\qquad
,\qquad  \nonumber \\
F_{tr}^{1} &=&-\partial _{r}({\cal V}^{-1}Y^{1})=-\frac{4q_{1}}{r^{3}\left(
1+\frac{q_{1}}{r^{2}}\right) ^{2}}.  \label{sol}
\end{eqnarray}

\bigskip To lift our solution to ten dimensions first we write
\[
Y^{0}={\cal V}^{\frac{1}{3}}X^{0},\quad Y^{1}={\cal V}^{\frac{1}{3}}X^{1},
\]
from which we deduce that
\[
\frac{X^{0}}{X^{1}}=\frac{Y^{0}}{Y^{1}}=\frac{1}{\sqrt{2}}e^{-4\hat{\phi}}=%
\frac{r^{2}}{2q^{0}}\left( 1+\frac{q_{1}}{r^{2}}\right) .
\]
The gauge fields are

\[
A_{t}^{0}=\frac{r^{2}}{q_{0}},\quad A_{t}^{1}=\frac{2}{1+\frac{q_{1}}{r^{2}}}%
.
\]

The ten dimensional metric is then
\begin{eqnarray*}
d\hat{s}^{2} &=&e^{\frac{5}{3}\hat{\phi}}\left( -{\cal {V}}%
^{-4/3}(1+g^{2}r^{2}{\cal V}^{2})dt^{2}+{\cal V}^{2/3}\left[ {\frac{dr^{2}}{%
1+g^{2}r^{2}{\cal V}^{2}}}+r^{2}(d\theta ^{2}+\sin ^{2}\theta d\phi
^{2}+\cos ^{2}\theta d\psi ^{2})\right] \right) \\
&&+e^{-\hat{\phi}}\left( \frac{q_{0}}{2r^{2}{\cal V}^{2}}dt^{2}\right) +e^{-%
\hat{\phi}}\left( \frac{\sqrt{2q_{0}}}{r{\cal V}}U_{m}^{1}\left( z\right)
dtdz^{m}\right) +e^{-\hat{\phi}}U_{m}^{a}\left( z\right) U_{n}^{a}\left(
z\right) dz^{m}dz^{n},
\end{eqnarray*}
where ${\cal V}=\frac{q_{0}^{\frac{1}{2}}}{4r}\left( 1+\frac{q_{1}}{r^{2}}%
\right) .$ The non vanishing components of the field strength of the
antisymmetric tensor $B_{MN}$ are
\begin{eqnarray*}
\hat{H}_{234} &=&e^{\frac{17}{6}\hat{\phi}}e_{0}^{t}e_{1}^{r}\partial
_{r}A_{t}^{0}=\frac{q_{0}^{\frac{7}{8}}}{2^{\frac{5}{16}}r^{\frac{7}{4}%
}\left( 1+\frac{q_{1}}{r^{2}}\right) ^{\frac{3}{8}}}, \\
\hat{H}_{015} &=&-\frac{1}{\sqrt{2}}e^{-\frac{7}{6}\hat{\phi}%
}e_{0}^{t}e_{1}^{r}\partial _{r}A_{t}^{1}=-\frac{2^{\frac{11}{16}}q_{1}}{%
q_{0}^{\frac{1}{8}}r^{\frac{11}{4}}\left( 1+\frac{q_{1}}{r^{2}}\right) ^{%
\frac{11}{8}}}, \\
\hat{H}_{567} &=&\frac{g}{2\sqrt{2}}.
\end{eqnarray*}

\subsection{\protect\bigskip Magnetic solutions}

\bigskip Here we will discuss the magnetic string solution found in \cite{m}%
. This is given by the metric
\begin{eqnarray}
ds^{2} &=&(gr)^{\frac{1}{2}}e^{-{\frac{3U}{2}}%
}(-dt^{2}+dz^{2})+e^{2U}dr^{2}+r^{2}\left( d\theta ^{2}+\sin ^{2}\theta
d\phi ^{2}\right)  \nonumber \\
e^{-U} &=&{\frac{1}{3gr}}+gr.
\end{eqnarray}
The gauge fields are given by $A_{\phi }^{I}=-q^{I}\cos \theta $ and the
scalar fields and the magnetic charges satisfy
\begin{equation}
X^{I}V_{I}=1,\qquad 3gq^{I}V_{I}=1.
\end{equation}

As for the magnetic solution of our model one finds that
\begin{equation}
X^{0}={\frac{1}{36}},\quad X_{0}=12,\quad X^{1}=3,\qquad X_{1}={\frac{2}{9}}%
,\quad e^{\frac{4\hat{\phi}}{3}}=3\sqrt{2}.
\end{equation}

The above solution can be lifted to ten dimensions and we get
\begin{eqnarray*}
d\hat{s}^{2} &=&\left( 18\right) ^{\frac{5}{8}}\left( (gr)^{\frac{1}{2}}e^{-{%
\frac{3U}{2}}}(-dt^{2}+dz^{2})+e^{2U}dr^{2}+r^{2}\left( d\theta ^{2}+\sin
^{2}\theta d\phi ^{2}\right) \right) \\
&&+2\left( 18\right) ^{-\frac{1}{8}}q_{1}^{2}\cos ^{2}\theta d\phi ^{2}-2^{%
\frac{3}{4}}\left( 18\right) ^{-\frac{1}{8}}q_{1}\cos \theta U_{m}^{1}\left(
z\right) d\phi dz^{m} \\
&&+\left( 18\right) ^{-\frac{1}{8}}U_{m}^{a}\left( z\right) U_{n}^{a}\left(
z\right) dz^{m}dz^{n}.
\end{eqnarray*}
\[
\hat{H}_{012}=e^{\frac{17}{6}\hat{\phi}}e_{3}^{\theta }e_{4}^{\phi }\partial
_{\theta }A_{\phi }^{0}=18^{\frac{17}{48}}\frac{q_{0}}{r^{2}}.
\]
We note that the numerical factors for the solution can be absorbed by
rescaling the coordinates and the charge $q_{1}.$ More solutions found in
\cite{SK} can also be lifted.

\bigskip

To lift these solutions further to eleven dimensions we have to first write
the dimensional reduction of eleven dimensional supergravity to $N=1$ ten
dimensional supergravity. These are \cite{chams1}
\[
E_{M}^{A}=e^{-\frac{1}{6}\hat{\phi}}\hat{e}_{M}^{A},\quad E_{\stackrel{.}{11}%
}^{11}=e^{\frac{4}{3}\hat{\phi}},\quad A_{MN\stackrel{.}{11}}=\hat{B}_{MN}
\]
and the eleven dimensional metric is related to the ten dimensional one by
\[
ds^{2\left( eleven\right) }=e^{-\frac{1}{3}\hat{\phi}}d\hat{s}^{2}+e^{\frac{8%
}{3}\hat{\phi}}\left( dx^{11}\right) ^{2}.
\]
The non vanishing components of the antisymmetric tensor field-strengths are
\[
F_{MNP\stackrel{.}{11}}=\hat{H}_{MNP}.
\]
The lifting of solutions from five to seven dimensions is very simple. We
write
\[
ds_{7}^{2}=e^{\frac{16}{15}\hat{\phi}}ds_{5}^{2}+e^{-\frac{8}{5}\hat{\phi}%
}\left( \left( dx^{5}\right) ^{2}+\left( dx^{6}\right) ^{2}\right) .
\]

\section{Conclusions}

In this work we have shown that it is possible to obtain $U(1)$ gauged $N=2$
five dimensional supergravity interacting with one vector multiplet by
compactifying and truncating ten dimensional supergravity on the group
manifold $SU(2)\times $ $U(1)^{2}$. The model obtained is the gauged version
of the supergravity model introduced by Strominger and Vafa. Using the
relation between the higher dimensional fields and the lower ones, it
becomes possible to lift known solutions such as black holes, string
solutions and domain walls of the five dimensional theory to seven, ten and
eleven dimensional supergravity theories to ten and eleven dimensions. Some
known electric and magnetic solutions for our gauged $D=5$ supergravity
compactified model, formulated in the framework of special geometry, were
lifted to higher dimensions. Such solutions are not easy to find directly by
studying the seven, ten and eleven dimensional supergravity theories. At
this stage it would be useful to study some of the properties of these
solutions and give their interpretation in terms of D-brane and M-theory
dynamics.


\begin{thebibliography}{99}
\bibitem{ads}  J.~M.~Maldacena, Adv.~Theor.~Math.~Phys.~{\bf 2} (1998) 231;
Int.~J.~Theor.~Phys.~{\bf 38} (1999) 1113;\newline
E.~Witten, Adv.~Theor.~Math.~Phys.~{\bf 2} (1998) 253;\newline
S.~S.~Gubser, I.~R.~Klebanov and A.~M.~Polyakov, Phys.~Lett.~{\bf B428}
(1998) 105.

\bibitem{gunaydin}  M. G\"{u}naydin, G. Sierrra, and P. K. Townsend, Nucl.
Phys.{\it \ }{\bf B253 }(1985) 573.

\bibitem{vaman}  H. Nastase and D. Vaman, \ {\it On the nonlinear KK
reductions on spheres of supergravity theories, }hep-th/0002028.

\bibitem{CV}  A. H. Chamseddine and M. S. Volkov, Phys. Rev. Lett. {\bf 79}
(1997) 3343; Phys. Rev. {\bf D57} (1998) 6242.

\bibitem{FS}  D. Z. Freedman, J. H. Schwarz, Nucl. Phys. {\bf B137} (1978)
333.

\bibitem{csd7}  A. H. Chamseddine and W. A. Sabra, $D=7$ $SU(2)$ {\it Gauged
Supergravity from }${\it D=}10$ {\it Supergravity}, hep-th/9911180.{\tt \ }%
To be published in Phys Lett. {\bf B.}

\bibitem{e12}  K. Behrndt, A. H. Chamseddine and W. A. Sabra, Phys. Lett.
{\bf B442 } (1998) 97.

\bibitem{e13}  K. Behrndt, M. Cvetic and W. A. Sabra, Nucl. Phys. {\bf B553}
(1999) 317.

\bibitem{m}  A. H. Chamseddine and W. A. Sabra, {\it Magnetic Strings in
Five Dimensional Gauged Supergravity Theories,} hep-th/9911195. To be
published in Phys Lett. {\bf B.}

\bibitem{SK}  D. Klemm and W. A. Sabra, \ {\it Supersymmetry of Black
Strings in D=5 Gauged Supergravities}, hep-th/0001131.

\bibitem{SV}  A. Strominger and C. Vafa, Phys. Lett. {\bf \ B379 }(1996) 99.

\bibitem{SS}  J. Scherk, J. H. Schwarz, Nucl. Phys. {\bf B153} (1979) 61.

\bibitem{townsend}  P.~S.~Howe and P.~K.~Townsend, {\em Supermembranes and
the modulus space of superstrings}, Talk given at Trieste Conference on
Supermembranes and Physics in $2+1$ Dimensions, Trieste, Italy, Jul 17 - 21,
1989, Published in Trieste Supermembr.~(1989) 165-172.

\bibitem{chams1}  A. H. Chamseddine, Nucl. Phys.{\it \ } {\bf B185} \ (1981)
403l ; Phys. Rev. {\bf D24} (1981) 3065 .
\end{thebibliography}
\end{document}